\documentclass[12pt]{iopart}
\usepackage{CJK}  
\usepackage{epsfig}  
\newcommand{\zc}{2.168~26(45)}
\newcommand{\dc}{0.057~650(12)}
\usepackage{iopams}  
\begin{document}
\begin{CJK*}{UTF8}{mj}

\title[Ising criticality of the IMD model]{Order-disorder transition in the two dimensional interacting monomer-dimer model: Ising criticality}
\author{Su-Chan Park (박수찬)}

\address{Department of Physics, The Catholic University of Korea, Bucheon 14662, Repulic of Korea}
\ead{spark0@catholic.ac.kr}
\vspace{10pt}
\begin{indented}
\item[]\today
\end{indented}

\begin{abstract}
We study the order-disorder transition of the two dimensional interacting 
monomer-dimer model (IMD) which has two symmetric absorbing states. To be self-contained,
we first estimate numerically the dynamic exponent $z$ of the two dimensional Ising model. From the relaxation dynamics of the magnetization at 
the critical point, we obtain $\beta/(\nu z) = \dc$, or $z = \zc$, where $\beta = \frac{1}{8}$ and $\nu=1$ 
are exactly known exponents. We, then, compare the critical relaxation of the order parameter at the transition 
point of the IMD with that of the Ising model. We found that the critical relaxation exponent $\beta/(\nu z)$ 
is in good agreement with the Ising model, unlike the recent claim by Nam \etal [JSTAT {\bf (2014)}, P08011].
We also claim that the Binder cumulant is not an efficient quantity to 
locate the order-disorder transition point of the model with two symmetric absorbing states.
\end{abstract}
\end{CJK*}
% Uncomment for PACS numbers
%\pacs{00.00, 20.00, 42.10}
%
% Uncomment for keywords
%\vspace{2pc}
%\noindent{\it Keywords}: XXXXXX, YYYYYYYY, ZZZZZZZZZ
%
% Uncomment for Submitted to journal title message
%\submitto{\JSTAT}
%
% Uncomment if a separate title page is required
\maketitle
% 
% For two-column output uncomment the next line and choose [10pt] rather than [12pt] in the \documentclass declaration
%\ioptwocol
%
\section{Introduction}
In two dimensions, systems with two symmetric absorbing states generally exhibit two phase 
transitions~\cite{DFL2003,ACDM2005,VL2008,CMP2009,NPKL2011,P2012}.
One is the symmetry breaking order-disorder transition (SBODT) and the other is the absorbing phase transition 
(APT). When these two transitions 
coincide~\cite{LV2011}, the critical relaxation dynamics shows logarithmic behavior like the voter 
model~\cite{Book:L1985} and, in this context, the universality class to which models with two symmetric 
absorbing states belong is termed as the generalized voter (GV) class~\cite{ACDM2005,DCCH2001,CCDD2005}. 

By capturing main feature of models with two symmetric absorbing states, Al Hammal \etal~\cite{ACDM2005} 
suggested a representative Langevin equation of the GV class, which resembles the model A (according
to the classification scheme of~\cite{HH1977}) with a single component. Because the critical
behavior of the model A with a single component order parameter is robust against various nonequilibrium 
perturbations~\cite{BS1994,GJY1985,TAS2002}\footnote{For the model A with multi-component order parameter,
certain nonequilibrium perturbations are relevant~\cite{DP2011}.}, the SBODT, once occurring at a distinct 
point from the APT point, is expected to share criticality with the Ising model. This indeed was 
confirmed numerically for the two-dimensional interacting monomers (2DIM) model~\cite{P2012}.

Recent numerical study of the two dimensional interacting monomer-dimer model (IMD) which also has two 
symmetric absorbing states, however, challenged this general picture and the SBODT of the IMD was claimed 
not to be the same as the Ising-type phase transition~\cite{NKL2014}. Nam \etal~\cite{NKL2014} 
argued that non-Ising criticality can be originated from interfacial fluctuation of the third state, 
so-called `dimer' state which is absent in the 2DIM. Hence more extensive numerical study
for the IMD is desired to settle down the issue of the universality class. The purpose of this paper is to clarify the universality 
class of the IMD by extensive Monte Carlo simulations.

This paper is organized as follows: In \sref{Sec:2}, the dynamic rules of the two dimensional IMD are 
explained.  \Sref{Sec:3} presents simulation results. Since the relaxation dynamics of the Ising model
at the critical point will play an important role in this paper, we present simulation results about the
dynamic exponent of the Ising model in \sref{Sec:Ising}. Then simulation results for the IMD are presented
in \sref{Sec:IMD}. \Sref{Sec:Sum} summarizes and concludes this work.

\section{\label{Sec:2}Two dimensional Interacting Monomer-Dimer Model}
As a variant of a catalytic surface reaction model proposed in~\cite{ZGB1986}, the IMD was first introduced  
as a one dimensional model with two symmetric
absorbing states~\cite{KP1994}  and the generalization to two dimensions was introduced and studied 
in~\cite{NPKL2011,NKL2014}. This section explains the dynamics of the two dimensional IMD through a 
simulation algorithm and introduces quantities we are interested in. Let us first denote the index of
each lattice point by $\vec{n}=(i,j)$ ($i,j = 1,2,\ldots, L$). 
Periodic boundary conditions are assumed. For convenience, a site $(i,j)$ will be called an even (odd) site if $i+j$ is even (odd).

Each site is one of three states; $A$-occupied, $B$-occupied, and vacant.
Each site $\vec{n}$ is assigned a random variable $\sigma_{\vec{n}}$ which takes one of 
three values $1$, $0$, and $-1$ if the site is occupied by $A$ ($\sigma_{\vec n}=1$), occupied by $B$ ($\sigma_{\vec n}=0$), or 
vacant ($\sigma_{\vec n}=-1$). By $A_{\vec{n}}$ [$B_{\vec{n}}$] is denoted
the number of $A$'s [$B$'s] in the nearest neighbors of site $\vec{n}$. 
If $\sigma_{\vec{n}} = -1$ (vacant) and $A_{\vec{n}}\neq 4$, the site $\vec{n}$ is referred to as 
an active site. If there is no active site, no change of configurations is allowed. In this sense,
a configuration without any active site is absorbing and there are two absorbing states;
all even sites are occupied by $A$ and all odd sites are vacant, and vice versa.\footnote{Actually, a
configuration in which all sites are occupied by particles does not allow
any further dynamics.
However, it is not considered an absorbing state, 
for no probability current to this configuration from any other state is present.}

Now we are ready for explaining an algorithm for simulations. Let us assume that there are $N_t$ active sites at time $t$. 
We choose one of active sites at random with equal probability. Assume that site $\vec n = (i,j)$ is selected.
With probability $p$ a `monomer' $A$ attempts to adsorb at this site (monomer-event) and with
probability $1-p$ a `dimer' $BB$ attempts to adsorb at site $\vec{n}$ and one of nearest neighbors of $\vec n$ (dimer-event).
For a dimer-event, the nearest neighbor site $\vec{m}$ can be either $(i+1,j)$ or $(i,j+1)$ with 
probability $\frac{1}{2}$. Note that even if we choose $\vec{m}$ among four
nearest neighbors with equal probability, the result is the same in the sense of probability.

The change by a monomer-event is as follows. If $A_{\vec n}= B_{\vec n}= 0$, that 
is, all nearest neighbors of site $\vec{n}$ are vacant, $A$ adsorbs at site $\vec{n}$, resulting in 
$\sigma_{\vec n}= 1$.
If $B_{\vec{n}}= 0$ but $A_{\vec{n}}\neq 0$, the adsorption attempt is neglected and no configuration change happens, which 
models strong repulsion between $A$'s. If $B_{\vec{n}}\neq 0$ , we choose 
one site among $B$-occupied nearest neighbors of $\vec{n}$ with equal probability. Assume that site $\vec{r}$ is chosen.
Then the adsorption-attempting $A$ and $B$ at site $\vec{r}$ react and form a `molecule' $AB$ 
which desorbs in no time.  
This event amounts to the change of $\sigma_{\vec{r}}$ from $0$ to $-1$.
In any case including dimer-events below, a `molecule' $AB$ 
formed by reaction is always removed from the system immediately.

A change by a dimer-event is as follows. If site $\vec{m}$ is not vacant,
the adsorption attempt is neglected and the configuration remains the same.
Assume that site $\vec{m}$ is also vacant. There are three cases which allow for a change of the configuration.
\begin{description}
\item[Case I:] $A_{\vec{n}} = A_{\vec{m}} = B_{\vec{n}} = B_{\vec{m}} = 0$.
\item[Case II:] $A_{\vec{n}} \neq 0$ and $A_{\vec{m}}=B_{\vec{m}}=0$; or $A_{\vec{m}} \neq 0$ and $A_{{\vec{n}}}=B_{{\vec{n}}}=0$.
\item[Case III:] $A_{\vec{n}} \neq 0$, $A_{\vec{m}} \neq 0$.
\end{description}
In all three cases, a dimer $BB$ is first dissociated into two
monomers which attempt to adsorb at site ${\vec n}$ and
${\vec m}$, individually.
In the case I, two $B$'s adsorb at sites $\vec{n}$ and $\vec{m}$, which gives $\sigma_{\vec{n}}=\sigma_{\vec{m}}=0$.
To explain what will happen in the case II, we let $\vec{s}$ [$\vec{r}$] be the site with nonzero [zero] number of $A$'s in its nearest neighbors.
Then, one $B$ adsorbs at site $\vec{r}$ and another $B$ reacts with one of $A$'s occupying nearest neighbors of  $\vec{s}$.
This event results in $\sigma_{\vec{r}}=0$ and $\sigma_{\vec{l}} = -1$, where $\vec{l}$ is a randomly chosen 
site among $A$-occupied nearest neighbors of $\vec{s}$.
In the case III, two $B$'s react individually with randomly chosen $A$'s. That is, the configuration becomes
$\sigma_{\vec{n}_n}= \sigma_{\vec{m}_n}=-1$, where $\vec{n}_n$ ($\vec{m}_n$) is a randomly chosen 
site among $A$-occupied nearest neighbors of $\vec{n}$ ($\vec{m}$).
Except these cases, no configuration change happens. 

After the adsorption attempt, regardless of whether it is successful or not, time increases by $1/N_t$.

For a given configuration, we introduce a random variable ${\cal M}$ to be called `staggered magnetization' (SM) as 
\begin{equation}
{\cal M} = \frac{1}{L^2} \sum_{i,j} (-1)^{i+j}\sigma_{i,j}.
\end{equation}
After many realizations with the same initial condition,
we calculate the mean SM, $m(t)$, and the (time-dependent) Binder cumulant, $U(t)$, defined as
\begin{equation}
m(t) = \left \langle {\cal M} \right \rangle_t,\quad
U(t) = 1 - \frac{\langle {\cal M}^4\rangle_t }{3 \langle {\cal M}^2 \rangle_t^2},
\end{equation}
where $\langle \ldots \rangle_t$ stands for the average at time $t$.
At the critical point of the SBODT, $m(t)$ decays to zero as $t^{-\beta/(\nu z)}$, where 
$\beta$ is the critical exponent for the order parameter, 
$\nu$ is the correlation length exponent, and $z$ is the dynamic exponent.
We will estimate $\beta/(\nu z)$ from simulations and compare it with that 
of the Ising model.

\section{\label{Sec:3}Simulation Results}
\subsection{\label{Sec:Ising}Preliminary : dynamic exponent of the two dimensional Ising model}
Since the main purpose of this paper is to figure out whether the SBODT of the IMD is described by the critical exponents of the Ising model,
we first present simulation results for the dynamic exponent of the Ising model at criticality, 
to be self-contained. This subsection may be read independently of other sections.
In this subsection, $\sigma$ and $m(t)$ should be understood as an Ising spin and the mean magnetization of 
the Ising model, respectively, and these should not be confused with the same notation for the IMD.

The Ising Hamiltonian is
\begin{equation}
H = - J \sum_{i,j=1}^L \sigma_{i,j} \left ( \sigma_{i+1,j}+ \sigma_{i,j+1}
\right ),
\end{equation}
where $\sigma_{i,j}$ is the Ising spin at site $(i,j)$ and 
periodic boundary conditions are assumed. 
We remind that the critical point and energy per site at the critical point are exactly known as 
$K_c \equiv J/k_B T_c = \ln (1+\sqrt{2})/2$ and $E_c/J = -\ln 2/2$~\cite{O1944}.
For convenience, we set $J=1$ and energy is measured in unit of $J$. 

We simulated the dynamics of the Ising model at the critical point, using single spin flip dynamics with the Metropolis algorithm. 
As an initial condition, a fully ordered state is used, that is, we set $\sigma_{i,j} = 1$ for all $i,j$ at $t=0$.
We measure  magnetization $m(t)$, fluctuation of magnetization $V(t)$, and
the energy per site $E(t)$, defined as
\begin{eqnarray}
m(t) &=& \left \langle \frac{1}{L^2}\sum_{i,j}  \sigma_{i,j} \right \rangle_t,\\
V(t) &=& \left \langle \left [ \frac{1}{L^2} \sum_{i,j}  \sigma_{i,j} \right ]^2\right \rangle_t
- m(t)^2,\\
E(t) &=& - \frac{1}{L^2} \left \langle \sum_{i,j} \sigma_{i,j} \left ( \sigma_{i+1,j}+ \sigma_{i,j+1} \right )
\right \rangle_t.
\end{eqnarray}
At the critical point, the asymptotic behaviors of these quantities are (see, for example, \cite{NKL2008})
\begin{eqnarray}
m(t) &=& A_m t^{-\beta / (\nu z)} \left [ 1 + B_m t^{-\chi_m}  + o(t^{-\chi_m}) \right ],\\
e(t) &\equiv& E(t) -E_c = A_e t^{-(\nu d -1)/(\nu z)} \left [ 1 + B_e t^{-\chi_e} + o(t^{-\chi_e}) \right ],\\
V(t) &=& A_v t^{(\nu d-2 \beta)/(z\nu)} \left [ 1 + B_v t^{-\chi_v} + o(t^{-\chi_v}) \right ],
\end{eqnarray}
where $d$ is the dimensions of the system ($d=2$ in this paper);
$\beta = \frac{1}{8}$ and $\nu = 1$ are exactly known critical exponents~(see, for example, \cite{Book:B1982}); $z$ is the
dynamic exponent to be determined; $A$'s and $B$'s are constants; and $\chi$'s are exponents of
the leading correction-to-scaling behavior which will be called leading corrections-to-scaling
exponents (LCEs). 
To find the dynamic exponent, we investigate the effective exponent functions (EEFs) $(b>1)$
\begin{eqnarray}
{\cal E}_m(t;b) &=& \frac{\ln[m(t)/m(t/b) ]}{\ln b} = -\frac{1}{8 z} - B_m \frac{b^{\chi_m} - 1}{\ln b} t^{-\chi_m}
+ o(t^{-\chi_m} ),\label{Eq:Em}\\
{\cal E}_e(t;b) &=& \frac{\ln[e(t)/e(t/b) ] }{\ln b} = -\frac{1}{z} - B_e \frac{b^{\chi_e} - 1}{\ln b} t^{-\chi_e} + o(t^{-\chi_e} ),\\
{\cal E}_v(t;b) &=& \frac{\ln[V(t)/V(t/b) ] }{\ln b} = -\frac{7}{4z} - B_v \frac{b^{\chi_v} - 1}{\ln b} t^{-\chi_v} + o(t^{-\chi_v} ).
\label{Eq:Ev}
\end{eqnarray}
Since the LCE governs how the EEF approaches the asymptotic value, 
it is also important to find the value of the LCE. 
To this end, we use the method introduced recently~\cite{P2013,P2014}.
We numerically calculate the corrections-to-scaling functions (CTSFs) $Q_m(t;b)$, $Q_e(t;b)$,
$Q_v(t;b)$, defined as
\begin{eqnarray}
Q_m(t;b) &=& \ln \left [ \frac{m(t)m(t/b^2)}{m(t/b)^2}\right ] = B_m (b^\chi_m - 1)^2 t^{-\chi_m} + o(t^{-\chi_m}),\label{Eq:Qm}\\
Q_e(t;b) &=& \ln \left [ \frac{e(t)e(t/b^2)}{e(t/b)^2}\right ]= B_e (b^\chi_e - 1)^2 t^{-\chi_e} + o(t^{-\chi_e}),\\
Q_v(t;b) &=& \ln \left [ \frac{V(t)V(t/b^2)}{V(t/b)^2}\right ]= B_v (b^\chi_v - 1)^2 t^{-\chi_v} + o(t^{-\chi_v}).
\end{eqnarray}
Note that the knowledge of critical indices is not necessary
to find $\chi$ from $Q$. 
For convenience, we just drop indices like $Q(t;b)$, ${\cal E}(t;b)$, $\chi$ 
when we have to write EEFs, CTSFs, or LCEs collectively in the following.

To estimate the critical exponents, we use the following strategy. 
First note that the correct value of $\chi$ makes $Q(t;b)/(b^\chi -1)^2$ for any $b$ lie on 
a single curve $B t^{-\chi}$ in the long time regime. Exploiting this feature,
we adjust the value of $\chi$ until double-logarithmic plots of   
$Q(t;b)/(b^\chi -1)^2$ as a function of $t$ for different $b$'s lie 
on a single straight line. 
After finding $\chi$, we plot ${\cal E}(t;b)$ as a function of 
$\tau \equiv (b^{\chi} - 1)t^{-\chi}/\ln b$ for various $b$'s. These curves should lie on a straight line
in the asymptotic regime, once $\chi$ is correctly estimated. 
By extrapolating the straight line behavior of ${\cal E}(t;b)$,
we can estimate the critical exponent.

\begin{figure}[t]
\centerline{
\includegraphics[width=0.6\textwidth]{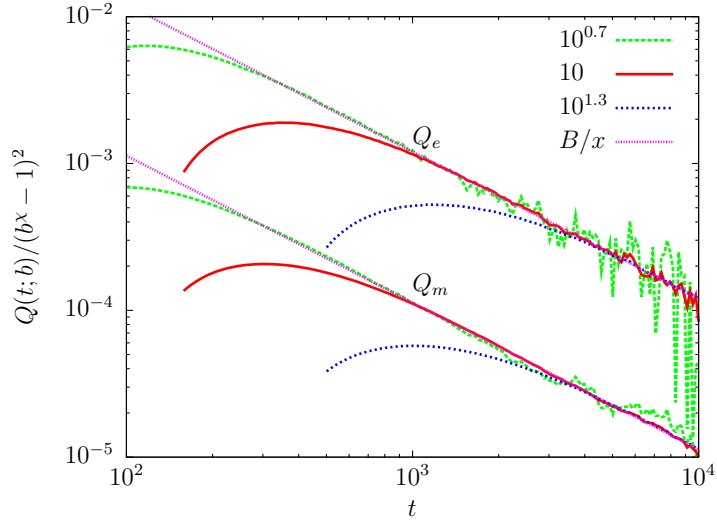}
}
\caption{\label{Fig:qme} Double-logarithmic plots of $Q(t;b)/(b^\chi-1)^2$ vs $t$ for
$b = 10^{0.7}\approx 5$, $b=10$, and $b=10^{1.3} \approx 20$  (left to right) with $\chi = 1$. 
Top three curves are for $Q_e(t;b)$ and
bottom three curves for $Q_m(t;b)$. Asymptotically, curves for different $b$'s lie
on a single curve, suggesting $\chi_m = \chi_e = 1$. 
For comparison, we also plot $B/x$ with $B$ obtained
from the fitting of the effective exponents; see \fref{Fig:I_eff} and \fref{Fig:E_eff}.
}
\end{figure}
\begin{figure}[t]
\centerline{
\includegraphics[width=0.6\textwidth]{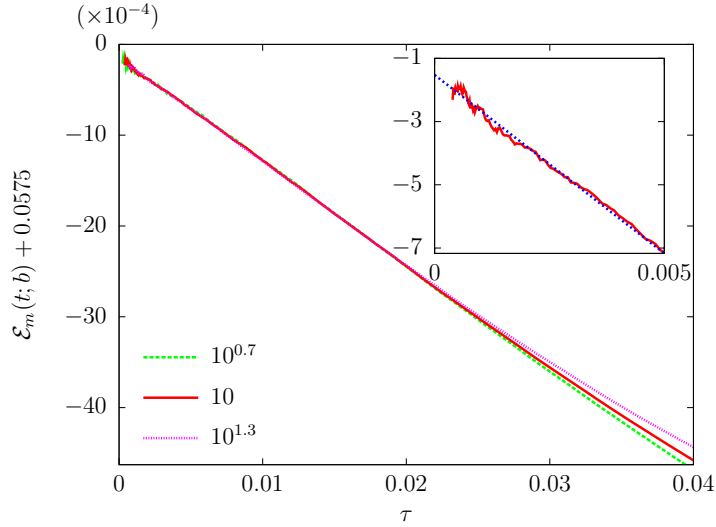}
}
\caption{\label{Fig:I_eff} Plots of ${\cal E}_m(t;b) + 0.0575$ vs $\tau \equiv (b-1)/( t \ln b)$ for
$b = 10^{0.7}$, 10, and $10^{1.3}$. Note that the 
scale on the $y$ axis is multiplied by $10^{-4}$. In the region $\tau \le 0.02$,
the effective exponents lie on a single straight line. 
The fitting for $\tau \le 0.015$ of the effective exponents gives $1/(8z) = 0.057~650(12)$ and $B_m = 0.114(2)$.
Inset: Close-up view on the region $\tau \le 0.005$ for $b=10$ with the 
fitted linear function (straight line).
}
\end{figure}
Now we present the simulation results.
In simulations, the system size is $L = 2^{12}$, the maximum observation time is $t=10^{4}$, and the number of independent
runs is $2\times 10^4$. We begin with the analysis of $Q_m$ and $Q_e$. 
\Fref{Fig:qme} depicts $Q(t;b)/(b^\chi-1)^2$ as functions of $t$ for a few $b$'s with
$\chi = 1$. Since the asymptotic behavior does not depend on $b$ with the choice of $\chi=1$, 
the LCEs are estimated as $\chi_m = \chi_e = 1$. 
Note that this estimate is consistent with the fitting result by Nam \etal~\cite{NKL2008}.

We now analyze the EEFs. \Fref{Fig:I_eff} shows how ${\cal E}_m(t;b)$ behaves
for sufficiently large $t$, when it is depicted against $\tau \equiv (b-1)/(t \ln b)$. 
Since EEFs are almost on the same line irrespective of $b$ for $\tau \le 0.02$, 
we conclude that the critical relaxation dynamics is well described by terms up to the leading correction for $\tau \le 0.02$. 
Fitting of ${\cal E}_m(t;10)$ for the region $\tau \le 0.015$
using a linear function, $1/(8z) - B_m \tau$ with $1/(8z)$ and $B_m$ to be fitting parameters, we
get 
\begin{equation}
\frac{1}{8z} = \dc,\quad B_m \approx 0.114(2),
\label{Eq:Ising_d}
\end{equation}
or equivalently
\begin{equation}
z = \zc,
\label{Eq:z}
\end{equation}
where the numbers in parentheses indicate the uncertainty of the last digits. We use
this convention throughout the paper.

\begin{figure}[t]
\centerline{
\includegraphics[width=0.6\textwidth]{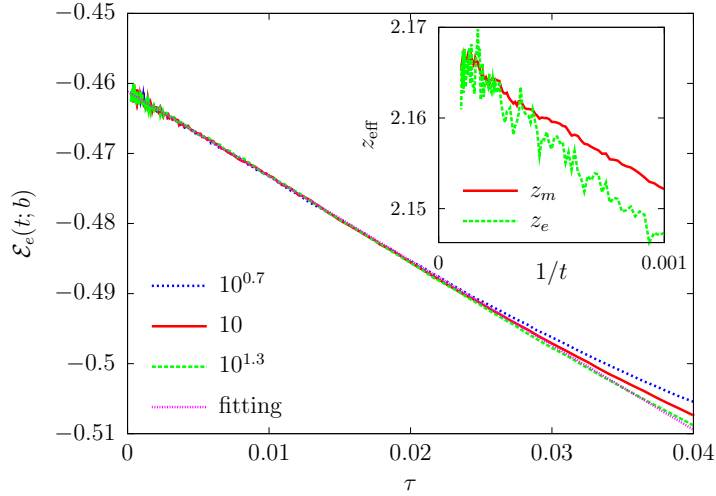}
}
\caption{\label{Fig:E_eff} Plots of ${\cal E}_e(t;b)$ vs $\tau \equiv (b-1)/( t \ln b)$ for
$b = 10^{0.7}$, 10, and $10^{1.3}$. The straight line is the result of the
fitting ${\cal E}_e(t;10)$ for the region $\tau \le 0.015$, using a linear fitting function.
Inset: Plots of $z_\textrm{\footnotesize eff}$ as a function of $1/t$ for the magnetization ($z_m$)
and the energy density ($z_e$).
}
\end{figure}
\begin{figure}[t]
\centerline{
\includegraphics[width=0.6\textwidth]{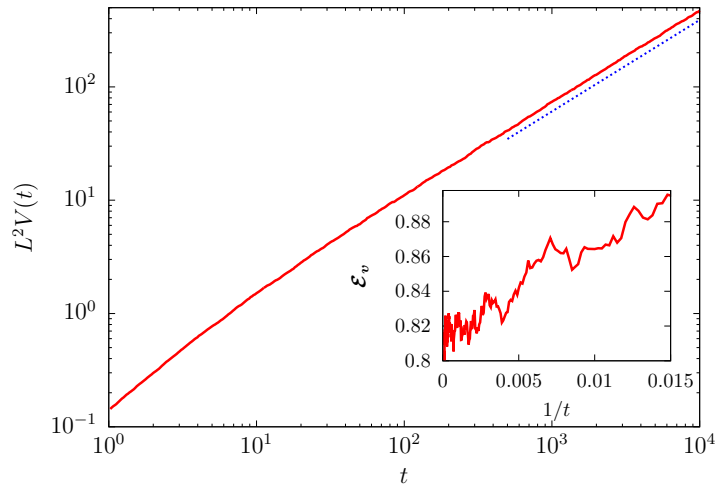}
}
\caption{\label{Fig:V_eff} Log-log plot of $L^2 V(t)$ vs $t$. A line segment with slope
$7/(4z) \approx 0.8071$ is for a guide to the eyes. 
Inset: Plot of ${\cal E}_v$ vs $1/t$ for $b=10$. 
}
\end{figure}
\Fref{Fig:E_eff} depicts the behavior of ${\cal E}_e(t;b)$ for some $b$'s. Similar to ${\cal E}_m$, all curves
for $\tau \le 0.02$ lie on  a single straight line. Due to the statistical noise
of the data, the estimate of $z$ is less accurate than \eref{Eq:z}. Still, we get a consistent
result within error bars. For comparison, we define effective exponents for $z$ as
\begin{eqnarray}
z_m(t) \equiv -\frac{1}{8 {\cal E}_m(t;10)},\quad
z_e(t) \equiv -\frac{1}{ {\cal E}_e(t;10)},
\end{eqnarray}
which are collectively called $z_\textrm{\footnotesize eff}$. The inset of \fref{Fig:E_eff} compares
these two effective exponents, which shows an agreement of the limiting values of two curves within errors.
 
\Fref{Fig:V_eff} depicts $L^2 V(t)$ as a function of $t$.
Since $V(t)$ is the fluctuation of $m(t)$, it is not surprising that $V(t)$ is much noisier than 
$m(t)$. The noisy behavior of $V(t)$ becomes conspicuous when ${\cal E}_v$ is depicted (see the inset
of \fref{Fig:V_eff}). Since ${\cal E}_v$ is quite noisy compared to ${\cal E}_m$ and ${\cal E}_e$, 
the error of the estimate of the leading 
scaling exponent, $7/(4z)$, solely from ${\cal E}_v$ is larger than before. From ${\cal E}_v$, 
we estimate $7/(4z) = 0.81(1)$ which is, of course, consistent with the estimate \eref{Eq:z} within errors.

As we have seen, statistical noise is minimal when the relaxation of magnetization is investigated.
Thus, we conclude $\beta/(\nu z) = \dc$ and, accordingly, $z = \zc$ for the single spin flip dynamics of the two dimensional Ising model with the Metropolis algorithm.

\subsection{\label{Sec:IMD}Interacting monomer-dimer model}
In this section, we present the simulation results for the IMD.
We begin with analyzing the EEF ${\cal E}_m(t)$ and the CTSF $Q_m(t)$ for the SM, defined similarly as \eref{Eq:Em} and \eref{Eq:Qm},
respectively. Since we do not know the critical point a priori, ${\cal E}_m(t)$ will be used to find the critical
point, by exploiting the fact that ${\cal E}_m(t)$ should veer up (down) as $t$ gets larger if the system is in the ordered (disordered) phase. 

\begin{figure}[t]
\centerline{
\includegraphics[width=0.6\textwidth]{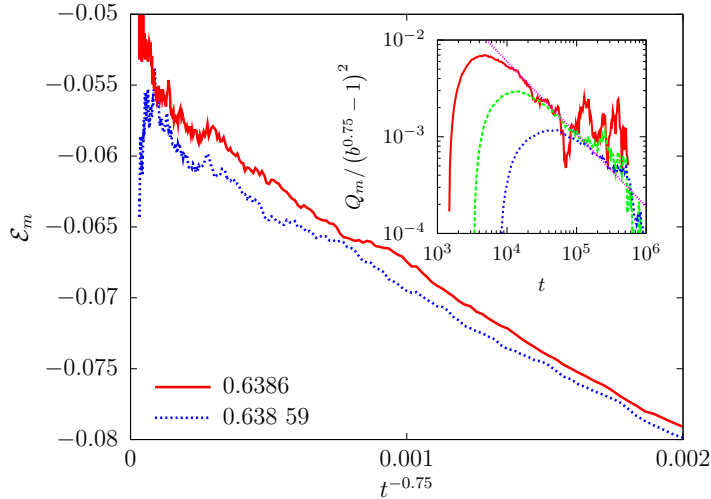}
}
\caption{\label{Fig:IMD_M} Plots of the effective exponents 
${\cal E}_m$ of the SM against  $t^{-0.75}$ for $p = 0.638~59$ and $0.6386$. 
Here, we are using $b=10$. Inset: Log-log plots of $Q_m/(b^\chi-1)^2$ vs
$t$ with $\chi = 0.75$ for $b = 10^{0.7}$, 10, $10^{1.3}$ (left to right).
The straight line with the slope 0.75 is also drawn for comparison.
}
\end{figure}

The initial condition of simulation is that  all odd sites are vacant and each 
even site can be either $A$-occupied with probability $m_0$ or vacant with probability $1-m_0$. With this initial preparation, the SM 
at time $t=0$ is $m(0) = m_0$. As in \cite{NKL2014}, we set $m_0  = 0.7$. \Fref{Fig:IMD_M} depicts ${\cal E}_m$ with $b=10$ as 
a function of $t^{-0.75}$ for $p = 0.638~59$ and $0.6386$. Here, the system size is $L = 2^{11}$, 
the number of independent runs is $1400$, and  the system evolves up to $t= 10^{6}$. The effective exponent for 
$p = 0.638~59$ [$0.6386$] approaches the Ising value~\eref{Eq:Ising_d} then veers down [up] for large $t$, suggesting that the critical point is $p_c = 0.638~595(5)$. 
This observation shows that the critical exponent of the IMD is consistent with the Ising value, contrary 
to the claim in \cite{NKL2014}.

The inset of \fref{Fig:IMD_M} shows the behavior of $Q_m/(b^\chi-1)^2$ at $p= 0.638~59$
with $\chi = 0.75$ for some values of $b$ on a double-logarithmic
scale. Since $Q_m$ is measured at the disordered phase, the curves should
eventually veer down~\cite{P2014} as can be seen at the tail of the curves
in the inset. Still, the power-law region is observable and we conclude 
that the CSE is about 0.75. Notice that the LCSE of the IMD is smaller than that of the Ising model (see \fref{Fig:qme}).
Thus, to find the correct value of critical exponents for the IMD requires longer evolution time than the Ising model,
which is also observed in~\cite{P2012} for the 2DIM.

\begin{figure}[t]
\centerline{
\includegraphics[width=0.6\textwidth]{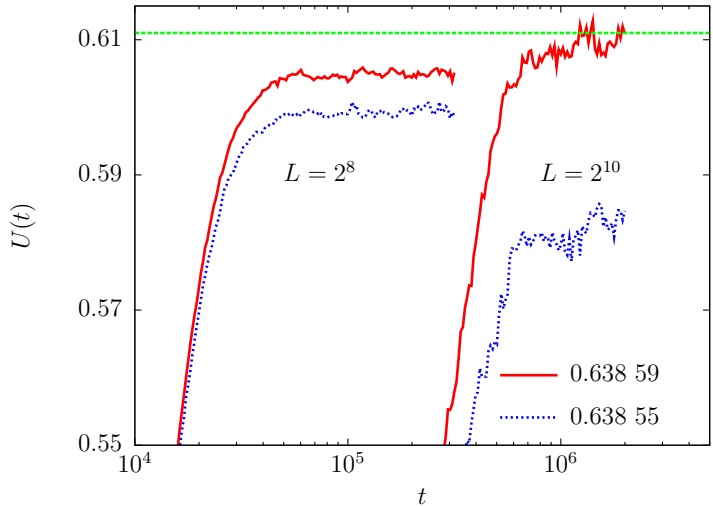}
}
\caption{\label{Fig:IMD_B} Plots of the time dependent Binder cumulant 
against $t$ on a semi-logarithmic scales for $L=2^8$ (left two
curves) and $L=2^{10}$ (right two curves) at $p= 0.638~55$ (top red curves)
and $p = 0.638~59$ (bottom blue curves). For comparison, a straight line
indicating the critical Binder cumulant of the Ising model (0.611)
is also drawn.
}
\end{figure}
The study of ${\cal E}_m$ suggests that the order-disorder transition of the IMD is indeed described by the critical 
exponents of the Ising model. To support our claim further, we also study the behavior of the Binder cumulant 
$U(t)$. As in \cite{NKL2014}, the initial condition for the study of the Binder cumulant is the fully vacant state without any $A$ 
and $B$. In \fref{Fig:IMD_B}, we showed the behavior of $U(t)$ at $p = 0.638~59$ (top red curves) and $p = 0.638~55$ (bottom
blue curves) for $L=2^8$ (left two curves) and $2^{10}$ (right two curves). 
The number of independent runs for $L=2^8$ and $p=0.638~55$ ($p=0.638~59$) is 40 000 (60 000)  and
the number of independent runs for $L=2^{10}$ and $p=0.638~55$ ($p=0.638~59$) is 2000 (3000).
At $p=0.638~55$ which is the estimated critical point in \cite{NKL2014}, we also observed that $U(t)$ approaches
around 0.6 for $L=2^{8}$ as in \cite{NKL2014}. However, $U(t)$ for $L= 2^{10}$ approaches 0.58, indicating that 
the system with $p = 0.638~55$ is in the disordered phase. This observation is consistent with our estimate of the critical point. 

As in the 2DIM studied in \cite{P2012}, analyzing the Binder cumulant is not an efficient method
to find the critical point of the SBODT for models with two symmetric absorbing states. Even though the system with $p = 0.638~59$ is 
in the disordered phase, the Binder cumulant 
increases up to $L = 2^{10}$, which might lead to a wrong conclusion. 
Since the Binder cumulant at $p=0.638~59$ is quite close to the Ising critical value for $L = 2^{10}$, 
to find the correct critical point using Binder
cumulant requires the system size to be larger than $L = 2^{10}$.

\section{\label{Sec:Sum}Summary and Conclusion}
We numerically studied the two-dimensional interacting monomer-dimer model (IMD), focusing on
the symmetry breaking order-disorder transition. Relaxation dynamics of the `staggered magnetization' around
the critical point was analyzed. 
The analysis of the corrections to scaling function showed that the two-dimensional
IMD model has stronger corrections to scaling than the Ising model. 
We found that the critical point of the IMD is $p_c = 0.638~595(5)$ and the relaxation dynamics at the 
critical point is consistent with the critical relaxation of the Ising model which is estimated as $\beta/(\nu z) = \dc$ in \sref{Sec:Ising}.
Thus, we concluded that the order-disorder transition in the two dimensional IMD shares criticality with the Ising 
model.
As a final remark, we would like to emphasize that due to strong corrections to scaling, the estimate of the critical point using the Binder cumulant becomes unreliable unless
the system size is larger than $2^{10}$.
\ack
The author thanks J. Krug for his hospitality during the author's visit to Universit\"at zu K\"oln in summer 2015, where 
this work was completed.  Financial support by the Basic Science Research Program through the
National Research Foundation of Korea~(NRF) funded by the Ministry of
Science, ICT and Future Planning~(No. 2014R1A1A2058694) is gratefully acknowledged.
Simulations were performed on the Cheops cluster at RRZK, K\"oln.
\section*{References}
%\References
%\begin{thebibliography}{99}
%\bibliographystyle{iopart-num}
\bibliographystyle{unsrt}
\bibliography{Park}
%\end{thebiliography}
\end{document}